\documentclass[article,nofootinbib]{revtex4-2}

\usepackage{amsmath}
\usepackage{graphicx}
\usepackage{epsf}
\usepackage{psfrag}
\usepackage{epsfig}
\usepackage{graphics}
\usepackage{amsfonts}
\usepackage{epstopdf}
\usepackage{slashed}
\usepackage{footnote}
\usepackage[brazilian]{babel}
\usepackage[utf8]{inputenc}
\usepackage[T1]{fontenc}
\usepackage[makeroom]{cancel}

\begin{document}

\newcommand{\be}{\begin{equation}}
\newcommand{\ee}{\end{equation}}
\newcommand{\bq}{\begin{eqnarray}}
\newcommand{\eq}{\end{eqnarray}}
\newcommand{\ba}{\begin{align}}
\newcommand{\ea}{\end{align}}

\newcommand{\Dslash}{\hbox{$\partial\!\!\!{\slash}$}}
\newcommand{\qslash}{\hbox{$q\!\!\!{\slash}$}}
\newcommand{\pslash}{\hbox{$p\!\!\!{\slash}$}}
\newcommand{\bslash}{\hbox{$b\!\!\!{\slash}$}}
\newcommand{\kslash}{\hbox{$k\!\!\!{\slash}$}}
\newcommand{\kbruto}{\hbox{$k \!\!\!{\slash}$}}
\newcommand{\pbruto}{\hbox{$p \!\!\!{\slash}$}}
\newcommand{\qbruto}{\hbox{$q \!\!\!{\slash}$}}
\newcommand{\lbruto}{\hbox{$l \!\!\!{\slash}$}}
\newcommand{\bbruto}{\hbox{$b \!\!\!{\slash}$}}
\newcommand{\parbruto}{\hbox{$\partial \!\!\!{\slash}$}}
\newcommand{\Abruto}{\hbox{$A \!\!\!{\slash}$}}
\newcommand{\bbbruto}{\hbox{$b_1 \!\!\!{\slash}$}}
\newcommand{\bbbbruto}{\hbox{$b_2 \!\!\!{\slash}$}}

\title{The full Lorentz-violating vacuum polarization tensor: low and high energy limits}

\date{\today}

\author{J. C. C. Felipe $^{(a)}$} \email[]{alexandre.vieira@uftm.edu.br}
\author{A. Yu. Petrov $^{(b)}$} \email[]{scarpelli@cefetmg.br}
\author{A. P. Ba\^eta Scarpelli $^{(c)}$} \email[]{petrov@fisica.ufpb.br}
\author{A. R. Vieira $^{(d)}$} \email[]{jean.cfelipe@ufvjm.edu.br}

\affiliation{(a) Instituto de Engenharia, Ci\^encia e Tecnologia, Universidade Federal dos Vales do Jequitinhonha e Mucuri, Avenida Um, 4050 - 39447-790 -Cidade Universit\'aria - Jana\'uba - MG - Brasil}
\affiliation{(b) Departamento de F\'isica, Universidade Federal da Para\'iba, Jo\~ao Pessoa - PB - Brasil}
\affiliation{(c) Centro Federal de Educa\c{c}\~ao Tecnol\'ogica - MG, Avenida Amazonas, 7675 - 30510-000 - Nova Gameleira - Belo Horizonte - MG - Brasil}
\affiliation{(d) Universidade Federal do Tri\^angulo Mineiro, Campus Iturama, Iturama - MG - Brasil}

\begin{abstract}
We compute the full vacuum polarization tensor in the fermion sector of Lorentz-violating QED. It turns out to be that even if we assume momentum routing invariance of the 
Feynman diagrams, it is not possible to fix all surface terms and find an ambiguity-free vacuum polarization tensor. The high and low energy limits of this tensor are obtained 
explicitly. In the high energy limit, only $c_{\mu\nu}$ coefficients contribute to the result. In the low energy limit, we find that Lorentz-violating induced terms  depend on 
$b_{\mu}$, $c_{\mu\nu}$ and $g_{\mu\nu\lambda}$ coefficients and vanish at $p=0$. At small $p$, we succeeded to obtain implications for condensed matter systems, explicitly, for 
the Hall effect in Weyl semi-metals. 

\end{abstract}


\maketitle

\section{Introduction}

Lorentz and CPT symmetries are known to be among the main criteria to formulate field theory models. To test how good these symmetries are, the Standard Model Extension (SME) \cite{Alan1} presents 
itself as the usual Standard Model extended by adding all possible Lorentz and CPT-violating terms. One possibility of the origin of these terms could be their emergence due to a spontaneous symmetry 
breaking that occurs at the Planck Scale \cite{Samuel}, explicit symmetry breaking or noncommutative field theory \cite{NonCom}. On the other hand, even if Lorentz and CPT symmetries are in fact exact 
at low energies, the question would be on what precision one can say that they are indeed valid. 

The tree level SME brings consequences to low energy physical models like quantum mechanical systems, and it can be used as a framework to test Lorentz and CPT symmetries in that
limit. By the way, most of the searches on Lorentz and CPT violation are based on non-relativistic Hamiltonians allowing us to see how SME coefficients affect usual quantum
mechanics. Some examples include spectroscopy \cite{Lamb} and condensed matter systems \cite{Grushin} (for other studies of Lorentz symmetry breaking within the condensed matter context see also f.e. \cite{FerrCM}). 

Beyond tree level, the minimal SME is one-loop  renormalizable, both in the electroweak \cite{Alan2,Colladay3} and the strong sectors \cite{Colladay,Colladay4}. There is also a recent 
investigation concerning Weyl semi-metals and terms induced by quantum corrections \cite{KLMSS}. However, results for finite quantum corrections coming from loop diagrams 
are usually controversial. There was a long standing debate concerning the issue of radiatively induced Carroll-Field-Jackiw (CFJ) terms \cite{CSI}-\cite{CSI11} suggesting
that loop computations are in general regularization dependent. The good old Dimensional Regularization \cite{DR, Bollini} can be used for computing the divergent 
part of the diagrams, as it was used in the proof of one-loop renormalizability. Unfortunately, it is not appropriate in some cases due to the presence of some Lorentz and 
CPT violating terms which contain objects which are well-defined only in specific dimensions, like Levi-Civita symbols and $\gamma_5$ matrices. In this case, computing the 
finite part of the amplitudes can be a delicate problem. The $\gamma_5$ issue can be avoided in certain situations \cite{Jegerlehner}, and there are also some recipes to treat it 
within traces involving Dirac matrices \cite{Cynolter, Tsai}. Besides this, the question what LV parameters contribute to perturbative corrections is interesting on 
its own.
 
In this work, we compute the full Lorentz-violating vacuum polarization tensor. We perform the computation of loop corrections within a four-dimensional implicit regularization
\cite{Nemes} framework, which does not assume any explicit regulator. The regularization-depended objects are mapped in surface terms. They manifest themselves as differences
between integrals with the same degree of divergence and so their value can yield any number including infinity. They are also the ones which can cause the breaking of symmetries 
of the model in a spurious way if explicitly computed. Therefore, we keep these terms intact till the end of the calculation and then require the fulfillment of a Ward-Takahashi 
or a Slavnov-Taylor identity. In this case, we guarantee gauge symmetry beyond tree level and at the same time find conditions on surface terms. As a consequence, not only
the induced CFJ term, but also other radiatively induced terms are arbitrary.

Another feature that can set values for surface terms is the momentum routing invariance (MRI) of the loop diagrams. For gauge field theories, there is a one-to-one
diagrammatic relation between gauge and MRI which is regularization independent. Some regularization schemes that are by construction momentum routing invariant, like dimensional
regularization, automatically fulfill gauge invariance. These conditions could be considered as another trial for finding equations that could fix arbitrary surface terms.
However, requiring MRI leads to the same relationships between the surface terms obtained when requiring gauge invariance and therefore at least one surface term remains
making the result arbitrary. All these reasoning on MRI do not cause any problem with the momentum routing diagrammatic computation of the chiral anomaly. Choosing the 
internal routing in order to fulfill the desired gauge Ward identity does not necessarily mean that momentum routing invariance was broken. Requiring MRI fixes the relation 
between surfaces terms and automatically fulfill gauge Ward-Takahashi identities and also reproduce the breaking of the axial current \cite{Vieira2}, the 
Adler-Bell-Jackiw anomaly. Although arbitrary and regularization dependent, the full Lorentz-violating vacuum polarization tensor involves only pieces proportional to $b_{\mu}$, 
$c_{\mu\nu}$ and $g_{\mu\nu\lambda}$ coefficients from the matter sector that affect the photon sector in the renormalization process.



The structure of the paper looks as follows: in section \ref{s1}, we list the relevant one-loop Feynman diagrams. In section \ref{s2},  we present the framework of the 
regularization. In section \ref{s3}, we calculate and present the quantum corrections. In section \ref{s5}, we discuss applications of our results to condensed matter. 
We present a summary in section \ref{s6} and a list of the relevant integrals in the Appendix.

\section{The framework and the one-loop diagrams}
\label{s1}

We consider the fermion sector of the standard minimal LV QED Lagrangian \cite{KosPic}:
\begin{align}
&\mathcal{L} =\frac{1}{2}i \bar{\psi}\Gamma^{\mu}\overleftrightarrow{D}_{\mu}\psi- \bar{\psi}M\psi-\frac{1}{4}F^{\mu\nu}F_{\mu\nu}-\frac{1}{4}(k_F)_{\kappa\lambda\mu\nu}F^{\mu\nu}F^{\kappa\lambda}+
\frac{1}{2}(k_{AF})^{\kappa}\epsilon_{\kappa\lambda\mu\nu}A^{\lambda}F^{\mu\nu},
\label{EQ1}
\end{align}
where $D_{\mu}\equiv \partial_{\mu}+iqA_{\mu}$ is the usual covariant derivative which couples the gauge field with matter,

\begin{align}
&\Gamma^{\nu}=\gamma^{\nu}+\Gamma^{\nu}_1,\nonumber\\
&\Gamma^{\nu}_1= c^{\mu\nu}\gamma_{\mu}+d^{\mu\nu}\gamma_5\gamma_{\mu}+e^{\nu}+i f^{\nu}\gamma_5+ \frac{1}{2}g^{\lambda\mu\nu}
\sigma_{\lambda\mu}
\end{align}
and
\begin{align}
&M=m+M_1, \nonumber\\
&M_1= i m_5 \gamma_5 +a^{\mu}\gamma_{\mu}+b_{\mu}\gamma_5\gamma^{\mu}+\frac{1}{2}H_{\mu\nu}\sigma^{\mu\nu}.
\end{align}

The coefficients $a_{\mu}$, $b_{\mu}$, $c_{\mu\nu}$, $d_{\mu\nu}$, $e_{\mu}$, $f_{\mu}$, $g_{\lambda\mu\nu}$, $H_{\mu\nu}$, $(k_F)_{\kappa\lambda\mu\nu}$ and
$(k_{AF})_{\kappa}$ break Lorentz symmetry and only the coefficients $a_{\mu}$, $b_{\mu}$, $e_{\mu}$, $f_{\mu}$, $g_{\lambda\mu\nu}$ and
$(k_{AF})_{\kappa}$ break CPT symmetry since their number of indices are odd. There is a particular subtleness in the $f_{\mu}$ coefficient because, unlike
 $a_{\mu}$ and $e_{\mu}$ ones, its time component is odd under all spacial reflections \cite{Bretf}.

\begin{figure}
 \includegraphics[trim=0mm 150mm 100mm 0mm, clip, scale=0.5]{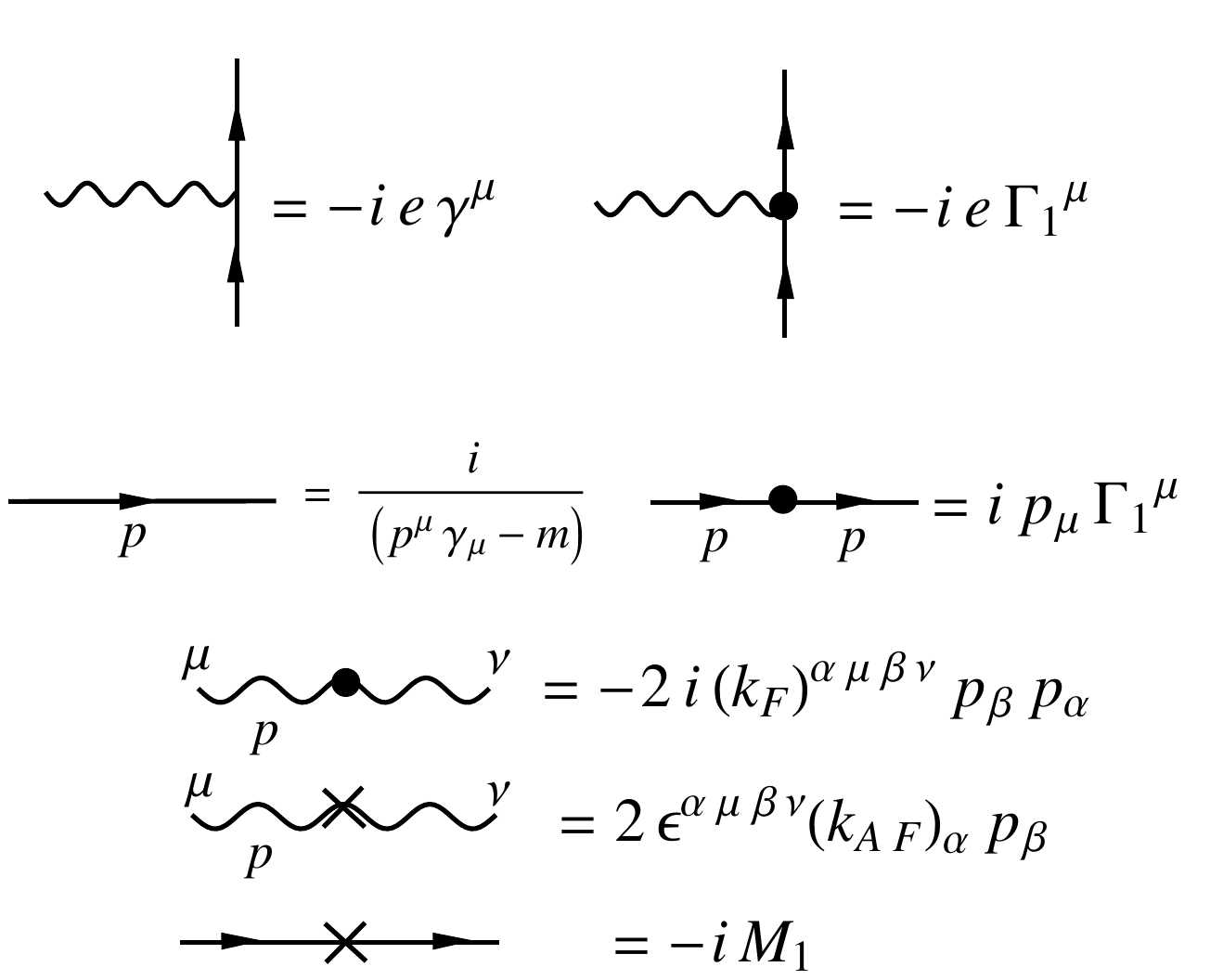}						
\caption{Feynman rules of the Lorentz-violating QED.}
\label{fig}
\end{figure}

The Feynman rules corresponding to the Lagrangian in eq. (\ref{EQ1}) are listed in Fig. \ref{fig}. We see that this Lagrangian involves a general vertex with
$\Gamma_\mu$ instead of just $\gamma_\mu$ and a general fermion propagator $\frac{i}{p_{\mu}\Gamma^\mu-M}$. However, considering this complete propagator for a generic $\Gamma^{\mu}$ is a difficult task. Therefore, the $\bullet$ and the $\times$ insertions in Fig. \ref{fig} denote the leading order in Lorentz and CPT violation in the fermion
propagator.

\begin{figure}
 \includegraphics[trim=0mm 60mm 0mm 0mm,scale=0.8]{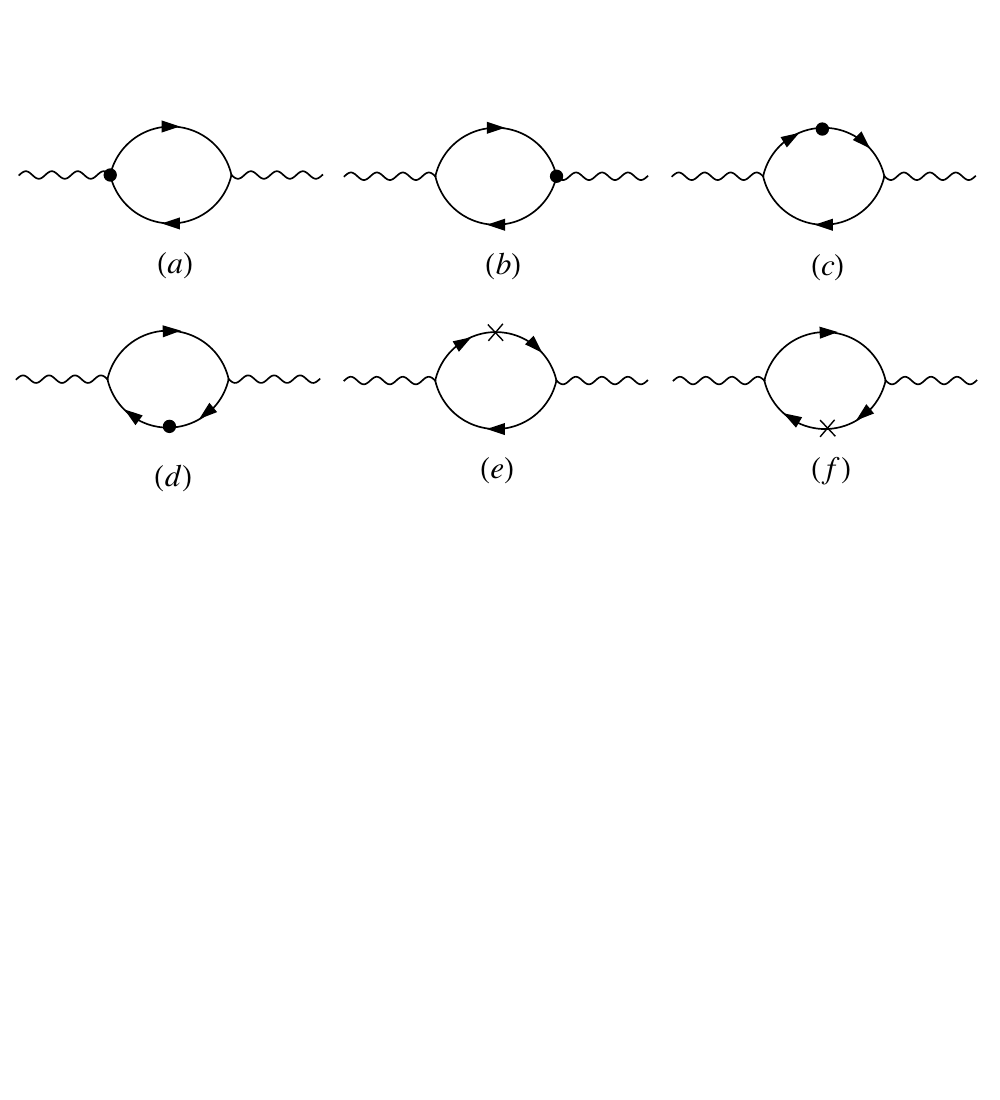}						
\caption{One-loop 2-point functions of the Lorentz-violating QED.}
\label{fig1}
\end{figure}

The one-loop diagrams are depicted in Fig. \ref{fig1}, and their amplitudes are:

\begin{widetext}
\be
\Pi^{\mu\nu}_{(a)}=-q^2\int \frac{d^4 k}{(2\pi)^4} Tr \left[\Gamma_1^{\nu}\frac{1}{\slashed{k}-m}\gamma^{\mu}\frac{1}{\slashed{k}-\slashed{p}-m}\right],
\label{(a)}
\ee

\be
\Pi^{\mu\nu}_{(b)}=-q^2\int \frac{d^4 k}{(2\pi)^4} Tr \left[\gamma^{\nu}\frac{1}{\slashed{k}-m}\Gamma_1^{\mu}\frac{1}{\slashed{k}-\slashed{p}-m}\right],
\label{(b)}
\ee

\be
\Pi^{\mu\nu}_{(c)}=q^2\int \frac{d^4 k}{(2\pi)^4} Tr \left[\gamma^{\nu}\frac{1}{\slashed{k}-m}\Gamma_1^{\lambda}k_{\lambda}\frac{1}{\slashed{k}-m}\gamma^{\mu}\frac{1}{\slashed{k}-\slashed{p}-m}\right],
\label{(c)}
\ee

\be
\Pi^{\mu\nu}_{(d)}=q^2\int \frac{d^4 k}{(2\pi)^4} Tr \left[\gamma^{\nu}\frac{1}{\slashed{k}-m}\gamma^{\mu}\frac{1}{\slashed{k}-\slashed{p}-m}\Gamma_1^{\lambda}
(k_{\lambda}-p_{\lambda})\frac{1}{\slashed{k}-\slashed{p}-m}\right],
\label{(d)}
\ee

\be
\Pi^{\mu\nu}_{(e)}=-q^2\int \frac{d^4 k}{(2\pi)^4} Tr \left[\gamma^{\nu} \frac{1}{\slashed{k}-m} M_1 \frac{1}{\slashed{k}-m} \gamma^{\mu} \frac{1}{\slashed{k}-
\slashed{p}-m} \right],
\label{(e)}
\ee

\be
\Pi^{\mu\nu}_{(f)}=-q^2\int \frac{d^4 k}{(2\pi)^4} Tr \left[\gamma^{\nu} \frac{1}{\slashed{k}-m} \gamma^{\mu} \frac{1}{\slashed{k}-
\slashed{p}-m} M_1 \frac{1}{\slashed{k}-\slashed{p}-m} \right].
\label{(f)}
\ee
\end{widetext}

Choosing of the regularization scheme to be applied is a subtle task because some LV terms involve objects defined only in the four-dimensional space-time, namely, the 
Levi-Civita symbol and $\gamma_5$ matrices. Thus, the inadequate choice of a regulator may cause spurious terms in these amplitudes and affect the conclusions below. On the
other hand, assuming that an implicit regulator exists allows us to manipulate the integrand and also to stay in $4$ dimensions and not to be concerned about spurious breaking terms 
in the process of renormalization.

\section{Basic Divergent Integrals and Surface Terms}
\label{s2}

Here we briefly describe the  framework  \cite{Nemes} (for a recent review see \cite{Carol}) allowing us to handle the divergent integrals in four dimensions which appear in the amplitudes of the previous section and establish some notation.  In this scheme, we assume that integrals are regularized by an
implicit regulator $\Lambda$ just to justify algebraic operations within the integrands. We then use, for instance, the following identity
\be
\int_k\frac{1}{(k+p)^2-m^2}=\int_k\frac{1}{k^2-m^2}-\int_k\frac{(p^2+2p\cdot k)}{(k^2-m^2)[(k+p)^2-m^2]},
\label{2.1}
\ee
where $\int_k\equiv\int^\Lambda\frac{d^4 k}{(2\pi)^4}$, in order to separate basic divergent integrals from the finite part. The former ones are defined as follows:
\begin{equation}
I^{\mu_1 \cdots \mu_{2n}}_{log}(m^2)\equiv \int_k \frac{k^{\mu_1}\cdots k^{\mu_{2n}}}{(k^2-m^2)^{2+n}}
\end{equation}
and
\begin{equation}
I^{\mu_1 \cdots \mu_{2n}}_{quad}(m^2)\equiv \int_k \frac{k^{\mu_1}\cdots k^{\mu_{2n}}}{(k^2-m^2)^{1+n}}.
\end{equation}

The basic divergences with Lorentz indices can be represented as differences between integrals with the same superficial degree
of divergence, according to the equations below, which define the surface terms \footnote{The Lorentz indices between brackets stand for permutations, {\it i.e.},
$A^{\{\alpha_1\cdots\alpha_n}B^{\beta_1\cdots\beta_n\}}=A^{\alpha_1\cdots\alpha_{n}}B^{\beta_1\cdots\beta_n}$
+ sum over permutations between the two sets of indices $\alpha_1\cdots\alpha_{n}$ and $\beta_1\cdots\beta_n$.}:
\begin{eqnarray}
&\Upsilon^{\mu \nu}_{2w}=  g^{\mu \nu}I_{2w}(m^2)-2(2-w)I^{\mu \nu}_{2w}(m^2)=\upsilon_{2w}g^{\mu \nu},
\label{dif1}\\
\nonumber\\
&\Xi^{\mu \nu \alpha \beta}_{2w}=  g^{\{ \mu \nu} g^{ \alpha \beta \}}I_{2w}(m^2)
 -4(3-w)(2-w)I^{\mu \nu \alpha \beta }_{2w}(m^2)=  \xi_{2w}g^{\{{\mu \nu}} g^{{\alpha \beta}\}},
\label{diff2}\\
\nonumber\\
&\Sigma^{\mu \nu \alpha \beta \gamma \delta}_{2w} = g^{\{\mu \nu} g^{ \alpha \beta} g^ {\gamma \delta \}}I_{2w}(m^2)
-8(4-w)(3-w)(2-w) I^{\mu \nu \alpha \beta \gamma \delta}_{2w}(m^2)= \sigma_{2w} g^{\{\mu \nu} g^{ \alpha \beta} g^ {\gamma \delta \}}.
\label{dif3}
\end{eqnarray}

In the expressions above, $2w$ is the degree of divergence of the integrals and  for the sake of brevity, we substitute the subscripts
$log$ and $quad$ by $0$ and $2$, respectively. Surface terms can be conveniently written as integrals of total
derivatives, namely
\begin{eqnarray}
\upsilon_{2w}g^{\mu \nu}= \int_k\frac{\partial}{\partial k_{\nu}}\frac{k^{\mu}}{(k^2-m^2)^{2-w}}, \nonumber \\
\label{ts1}
\end{eqnarray}
\begin{eqnarray}
(\xi_{2w}-v_{2w})g^{\{{\mu \nu}} g^{{\alpha \beta}\}}= \int_k\frac{\partial}{\partial
k_{\nu}}\frac{2(2-w)k^{\mu} k^{\alpha} k^{
\beta}}{(k^2-m^2)^{3-w}}.
\label{ts2}
\end{eqnarray}
and
\begin{eqnarray}
(\sigma_{2w}-\xi_{2w})g^{\{ \mu \nu} g^{ \alpha \beta} g^ {\gamma \delta \}} =
\int_k\frac{\partial}{\partial k_{\nu}}\frac{4(3-w)(2-w)k^{\mu} k^{\alpha} k^{\beta} k^{\gamma} k^{\delta}}{(k^2-m^2)^{4-w}}.
\label{ts3}
\end{eqnarray}

Equations (\ref{dif1})-(\ref{dif3}) are in principle arbitrary and regularization dependent. From the mathematical point of view, a surface term
can be any number since it is a difference between two infinities. They can be shown to vanish in usual dimensional regularization and they can
be finite or infinite if computed with a sharp cutoff. In general, we leave these terms unevaluated until the end of the calculation to be fixed on symmetry
grounds or phenomenology, when it is the case \cite{Jackiw}.

To illustrate this method, it is instructive to consider first the usual vacuum polarization tensor \cite{Felippe}:
\bq
&\Pi^{\mu\nu}(p)=\frac{4}{3}(p^2\eta^{\mu\nu}-p^{\mu}p^{\nu})I_{log}(m^2)-4\upsilon_2 \eta^{\mu\nu}+\frac{4}{3}(p^2\eta^{\mu\nu}-p^{\mu}p^{\nu})\upsilon_0 - \nonumber\\
&-\frac{4}{3}(p^2\eta^{\mu\nu}+2p^{\mu}p^{\nu})(\xi_0 -2\upsilon_0 ) -\frac{8i}{(4\pi )^2}(p^2\eta^{\mu\nu}-p^{\mu}p^{\nu})\int^1_0 x(1-x) \ln \frac{m^2-p^2 x(1-x)}{m^2},
\label{vptqed}
\eq
where $\upsilon_0$ and $\xi_0$ are logarithmic surface terms. Note that if we use a Ward identity, the surface term $\upsilon_2$ is fixed to be equal to zero, and we get a relation between
$\xi_0$ and $\upsilon_0$. At the same time, it is not possible to fix the $\upsilon_0$ term because this term is already gauge invariant. This surface term is the just same that appears 
in the CFJ induced term. In that case it is not possible to fix it as well because it is proportional to a Levi-Civita symbol.

The use of eq. (\ref{2.1}) is not the only one possible since the implicit regulator assumed allows any other operation with the integrands. It is, however, the equation we use 
the most in order to separate divergent from the finite part, because the second term in the r.h.s. of this equation is less divergent than the first. Disadvantages of 
doing this include the high number of powers in $k$ that can appear in the numerator of the integrals, which makes them difficult to compute (by the way, one manner to do this is 
proposed in \cite{Felippe}), and the surface terms that cannot be completely fixed by symmetries if only eq. (\ref {2.1}) is used.

\section{Evaluation of the diagrams}
\label{s3}

After taking the traces in eqs. (\ref{(a)})-(\ref{(f)}), regularizing the integrals and summing all the diagrams we find the full one-loop vacuum polarization
tensor. We list all integrals in the appendix.

\bq
&& \Pi^{\mu \nu}_{LV}=\frac 83 q^2 \left\{\left(c^{\mu \alpha}p^\nu+ c^{\nu \alpha} p^\mu\right)p_\alpha
- c^{\alpha \beta}p_\alpha p_\beta \eta^{\mu \nu} - p^2 c^{\mu \nu}\right\}
\left\{I_{log}(m^2)-\frac{i}{16 \pi^2}\left[\frac{(p^2+2m^2)}{p^2} Z_0 + \frac 13\right]+\frac{\upsilon_0}{2}\right\} +\nonumber \\
&& + \frac{i}{16 \pi^2} q^2 c^{\alpha \beta}p_\alpha p_\beta \frac {1}{p^2} \left(p^\mu p^\nu - p^2 \eta^{\mu \nu}\right)
\left\{p^2 \iota_0+ \frac{(p^2+4m^2)}{p^2}Z_0+\frac 23\right\}+ \nonumber \\
&& -\frac {mq^2}{2\pi^2}p_\lambda\left\{p^2 g^{\mu \nu \lambda}
+ p_\beta\left(g^{\nu \beta \lambda}p^\mu - g^{\mu \beta\lambda}p^\nu\right)\right\}\iota_1
-4imq^2 p_\alpha \left(g^{\nu \mu \alpha} - g^{\alpha \mu \nu} + g^{\alpha \nu \mu}  \right)\upsilon_0 + \nonumber \\
&& + q^2\left(- \frac{m^2}{\pi^2} \iota_0 +4 i \upsilon_0\right)b_\alpha p_\beta \epsilon^{\alpha \beta \mu \nu},
\label{result}
\eq
in which $Z_n=\int_0^1 dx\ x^n \ln \left[\frac{m^2-p^2 x(1-x)}{m^2}\right]$ and $\iota_n=\int_0^1 dx \frac{x^n(1-x)}{m^2-p^2 x(1-x)}$. Besides, we note that $c^{\mu\nu}$ is symmetric, and $g^{\mu\nu\alpha}$ is antisymmetric in the two first indices, as expected, since only these parts of the above mentioned tensors can contribute to observables. In order to obtain the result given by eq. (\ref{result}), the following relations involving the integrals in the Feynman parameters were used:
\be
Z_k=\frac{1}{k+1}\left\{k Z_{k-1} - (k-1)\frac{m^2}{p^2}Z_{k-2}-\frac{k-1}{k(k+1)}\right\},
\ee
\be
\iota_{k+1}=\frac 12 \left\{\iota_k-\frac {1}{p^2}\left[k Z_{k-1}-(k+1)Z_k\right]\right\}.
\ee

In eq. (\ref{result}), we have constrained the surface terms such that the result in eq. (\ref{result}) is transverse. This procedure fixes $\xi_0=2 \upsilon_0$, $\sigma_0=3 \upsilon_0$ and $\upsilon_2=\xi_2=0$. With these relations between the surface terms, the contributions proportional to the vectors $a^\mu$ and $e^\mu$ turned out to be zero. We see that gauge invariance of the action is not sufficient to determine all the surface terms, which implies the ambiguity of the induced CFJ term \cite{Carroll}. It is important to notice that the contributions due to the parameter $\upsilon_0$ in the terms with the tensors $c^{\mu \nu}$ and $g^{\mu \nu \alpha}$ are irrelevant since they can be eliminated through renormalization or by imposing some normalization condition.

One could try to determine $\upsilon_0$ by enforcing momentum routing invariance in the diagrams. One possible procedure would be to calculate the amplitude with arbitrary routing, parameterized by an alpha constant, and then require the result to not depend on such a parameter. It can be shown, for a QED amplitude $T^{\mu_1 \mu_2 \cdots \mu_n}$, with $n$ external photon legs, that its transversality only can be respected if a relative shift is allowed between the remaining $(n-1)$-point functions which result from the contraction of the external momentum $p_{\mu_i}$ with $T^{\mu_1 \mu_2 \cdots \mu_n}$. This relative shift is only allowed if some relation between surface terms are established. Here, we show this for $n=2$.

\begin{figure}[!h]
\centering
\includegraphics[trim=0mm 65mm 0mm 50mm, scale=0.25]{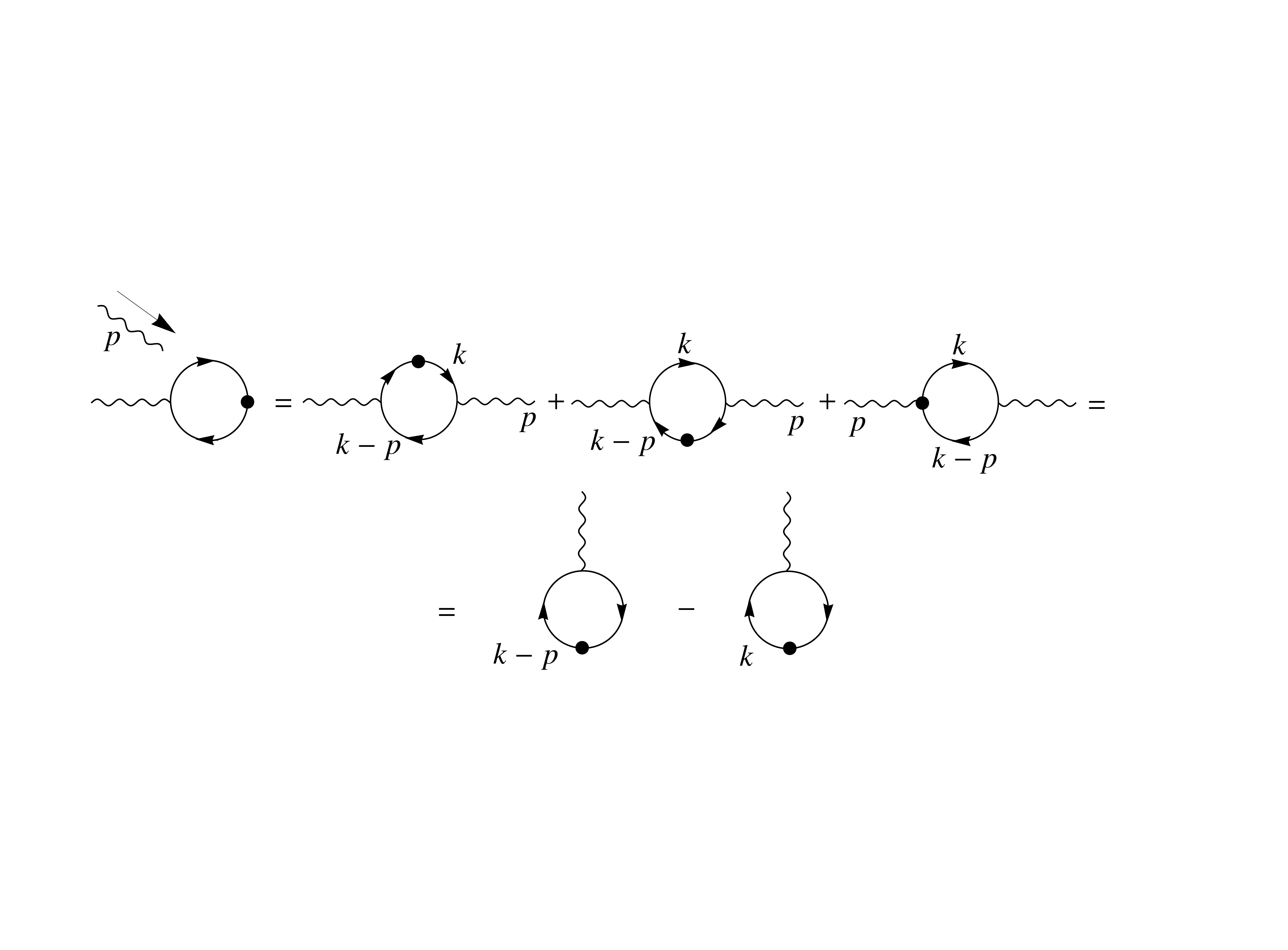}
\caption{Gauge and momentum routing invariance relation for a two leg diagram.}
\label{fig5}
\end{figure}

\begin{figure}[!h]
\centering
 \includegraphics[clip, trim=20mm 150mm 70mm 130mm, scale=0.25]{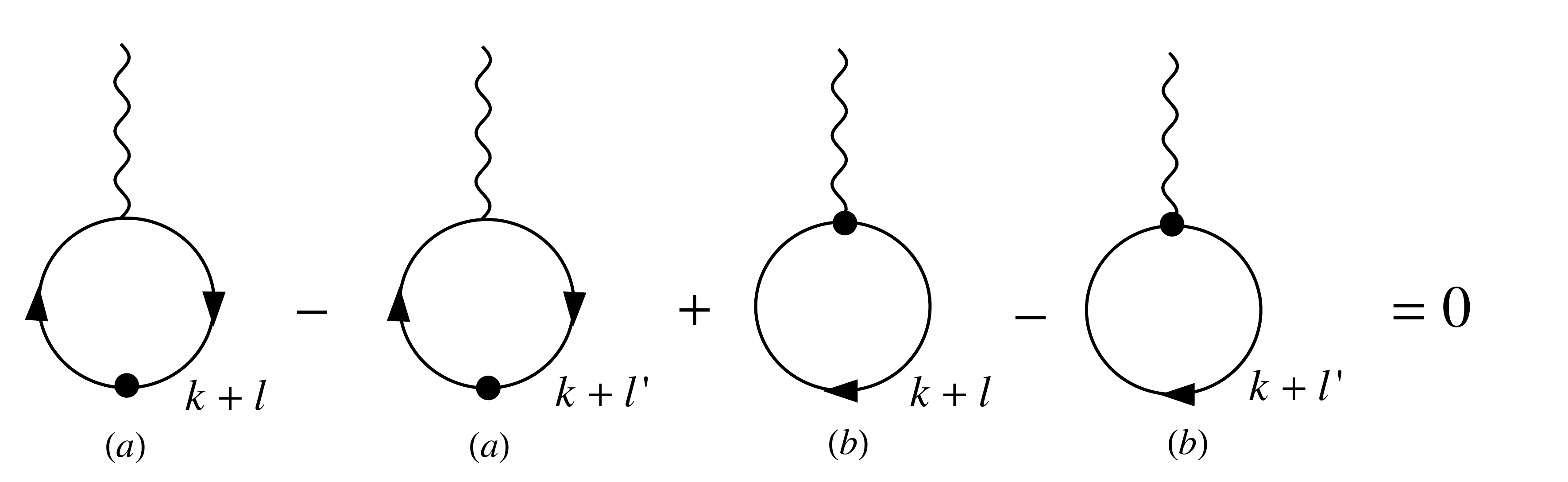}						
\caption{Momentum routing invariance of a tadpole.}
\label{figTad}
\end{figure}

We attribute a general loop momentum in the diagrams of Fig. \ref{fig5}, respecting the energy-momentum conservation in the vertices. When the external momentum $p_\mu$ is contracted with the diagrams, each one of the graphs, after the contraction is decomposed in a difference of two identical tadpole diagrams with different loop momentum. Considering the six two-point amplitudes, after the contraction with $p_\mu$, only four tadpole diagrams survive, which are shown in Fig. \ref{figTad}, in which $l$ is an general routing that is proportional to the external momentum, {\it i.e.}, $l=\alpha p$. The tadpoles are functions of $l$ ($l'$), $\tau ^{\nu}(l)$. The result of the calculation represented by Fig. \ref{figTad} is given by:

\begin{align}
&\tau ^{\nu}_{(a)}(l)-\tau ^{\nu}_{(a)}(l')+\tau ^{\nu}_{(b)}(l)-\tau ^{\nu}_{(b)}(l')=
8 q c_{\mu \alpha}\left\{(\alpha^3-\alpha'^3)\left[p^\mu p^\alpha p^\nu +p^2p^\alpha \eta^{\mu \nu}\right](-v_0+2\xi_0-\sigma_0)+ \right. \nonumber \\
& \left.+p^\alpha \eta^{\mu \nu}(2\upsilon_2-\xi_2)\right\}
+4q(\alpha^2-\alpha'^2)(p^2 m e^{\nu}+2 m p^{\nu}e\cdot p)(2\upsilon_0-\xi_0).
\label{Tad}
\end{align}

Since $\alpha\neq\alpha'$ by definition, the only possible solution for preserving gauge invariance (the tranversality of the photon polarization tensor) is the same that 
assures momentum routing invariance. In fact, the explicit calculation which results in eq. (\ref{Tad}) enforces that $\xi_0=2\upsilon_0$ and $\sigma_0=3\upsilon_0$. If we take 
into account the contributions from the traditional QED, we also obtain that $\upsilon_2=\xi_2=0$. It is interesting to remark that, although we have considered an arbitrary 
loop momentum in the vacuum polarization tensor, this result implies momentum routing invariance of the tadpole diagram, which results in transversality of the photon two-point 
function. In general, the transversality of an amplitude with $n$ external legs will result in routing independence of an amplitude with $n-1$ outer legs. This condition is weaker 
than the independence of loop-momentum of the original amplitude. Also, MRI looks like a symmetry of the Feynman diagrams, as in Fig. \ref{figTad}. But is not a symmetry in 
terms of an action, {\it i. e.} transformations of the fields that make this action invariant.

The piece $I_{log}(m^2)$ is a logarithmically divergent integral, namely $\int \frac{d^4k}{(2\pi)^4}\frac{1}{(k^2-m^2)^2}$, introduced in
section \ref{s2}. We do not have to evaluate it explicitly. It gives rise to $1/\epsilon+...$ in dimensional regularization, for example.
We now take the low energy limit ($m^2>>p^2$) in each term and integral of eq. (\ref{result}). For instance, $Z_1(m^2>>p^2)\approx -\frac{p^2}{12m^2}$ in this
limit. Thus, we find the low energy limit of the vacuum polarization tensor ($m^2>>p^2$):
\begin{widetext}
\begin{align}
& \Pi^{\mu \nu}_{LV}(p)=\frac 83 q^2 \left\{c^{\mu p}p^\nu+ c^{\nu p} p^\mu
- c^{pp}\eta^{\mu \nu} - p^2 c^{\mu \nu}\right\}\left( I_{log}(m^2)+ \frac{i}{16\pi^2} \frac{p^2}{6m^2} + \frac{\upsilon_0}{2}\right) + \nonumber\\
&+\frac{i}{16\pi^2}\frac{q^2}{3m^2} c^{pp}\left( p^{\mu}p^{\nu}-p^2\eta^{\mu\nu} \right)-\frac{mq^2}{12\pi^2}\frac{p^2}{m^2}
\left\{ g^{\mu\nu p}+\frac{1}{p^2}(p^{\mu}g^{\nu p p}-p^{\nu}g^{\mu p p})\right\}+ \nonumber\\
& -4imq^2 p_\alpha \left(g^{\nu \mu \alpha} - g^{\alpha \mu \nu} + g^{\alpha \nu \mu}  \right)\upsilon_0
- \left(\frac{q^2}{2\pi^2}\right) \epsilon^{bp\mu\nu}\left( 1- 8i\pi^2 \upsilon_0\right) \nonumber \\
&\approx   \frac 83 q^2 \left\{c^{\mu p}p^\nu+ c^{\nu p} p^\mu
- c^{pp}\eta^{\mu \nu} - p^2 c^{\mu \nu}\right\} I_{log}(m^2) +\frac{i}{16\pi^2}\frac{q^2}{3m^2} c^{pp}\left( p^{\mu}p^{\nu}-p^2\eta^{\mu\nu} \right)+\nonumber\\
&- \left(\frac{q^2}{2\pi^2}\right) \epsilon^{bp\mu\nu}\left( 1- 8i\pi^2 \upsilon_0\right),
\label{Pi1}
\end{align}
\end{widetext}
where $c^{p\nu}$ stands for $c^{\mu\nu}p_{\mu}$.

The result in eq. (\ref{Pi1}) shows that only $b_{\mu}$, $c_{\mu\nu}$ and $g_{\mu\nu\lambda}$ affect usual QED at low energies. The one-loop LV contribution to spectroscopy
or condensed matter physics would be due only to these terms. In particular, the radiatively induced CFJ term
$\epsilon^{\alpha\beta\mu\nu}b_{\alpha}p_{\beta}$ is the one most studied within various contexts and it is dominant compared to terms proportional to $g^{\mu\nu\lambda}$.

On the other hand, in the high-energy limit ($m^2<<p^2$) we find that vacuum polarization tensor is affected only by $c_{\mu\nu}$ coefficients:

\bq
&&\Pi^{\mu \nu}_{LV}(p) = \frac 83 q^2 \left\{c^{\mu p}p^\nu+ c^{\nu p} p^\mu
- c^{pp}\eta^{\mu \nu} - p^2 c^{\mu \nu}\right\}\left\{ I_{log}(m^2)- \frac{i}{16\pi^2} \left[ \ln{\left(-\frac{p^2}{m^2} \right)} -\frac 53 \right] + \frac{\upsilon_0}{2}\right\}+ \nonumber\\
&& -\frac{i}{16\pi^2}\frac 43 q^2 c^{pp} \left( \frac{p^\mu p^\nu}{p^2} - \eta^{\mu\nu} \right),
\label{pihigh}
\eq
where pieces proportional to $p_{\mu}$ can be disregarded since they couple with currents, and contributions proportional to $\partial_{\mu}J^{\mu}$ vanish due to the gauge 
symmetry.

\section{The parameter $b_{\mu}$ and applications to condensed matter}
\label{s5}

The bridge between high-energy physics and condensed matter has shown promising in recent years and the question of studies on the renormalization group has been the great trump card of this relationship \cite{Amit2005}. With the emergence of graphene, low-dimensional systems could be described via the massless Dirac equation in ($2+1$)-dimensional space-time. The application of field theory in low-dimensional systems has become a reality, with regard to theoretical studies, such as the model that describes how electrons propagate over a sheet of graphene, from the point of view of the renormalization group \cite{Gonzalez1994}. Another interesting application consists in considering curved space effects in graphene \cite{Vozmediano2010}, showing how useful is the application of field theory techniques to low-dimensional electronic systems.

The proposal that particles presenting a relativistic scattering relationship can be considered as quasi-particles in condensed matter models has been known for some time \cite{Gongolin1998}. In some models, the dispersion relation can be linearized, by means of an expansion around the Fermi energy, which ends up with relativistic energy-momentum relations. In this sense, Dirac's fermions gained some prominence with the discovery of graphene which, being an essentially ($2+1$)-dimensional system, is nothing more than a sheet composed of carbon atoms in a hexagonal lattice \cite{CastroNeto2009}. Thus, electrons moving over this sheet of carbon interact with the potential of this lattice, giving rise to conical structures close to Fermi energy. Near this region (called a node), the scattering relationships of electrons on the graphene sheet turn out to be linear and the Hamiltonian that describes the system is given by the non-massive Dirac equation, where the propagation velocity is the Fermi velocity ($v_{f}$) and the spin of the particle gives rise to a pseudo-spin, which is related to the sub-lattice of the system.

After the observation of Dirac fermions, it was realized \cite{Xu2011} that some materials whose band structure present nodes around the excitation of the material are Weyl fermions, such materials being called Weyl semi-metals \cite{Turner2013}. Initially, the Weyl semi-metals proposals were based on studies of pyrochlore iridium molecules, topological insulators and heterostructures \cite{Burkov2011}. Such descriptions paved the way for the distance between condensed matter and high energy physics to become smaller with regard to the description of certain phenomena, more specifically the emergence of field theory according to Anti-de Sitter-conformal field theory (AdS-CFT)  correspondence or Anti-de Sitter-condensed matter theory correspondence (AdS-CMT) \cite{Sachdev2013}. In the case of Weyl semi-metals, this connection occurs through the chiral anomaly, which can be translated into condensed matter models. The chiral anomaly basically tells us that the conservation laws of the vector current ($\partial_{\mu}j^{\mu}=0$) and that of the chiral current ($\partial_{\mu}j^{\mu}_{5}=0$) cannot be satisfied at the same time. Therefore, if we enforce vector current conservation, the chiral current cannot be conserved, which leads to the well-known chiral symmetry breaking. From the point of view of Weyl semi-metals, it can be written in terms of the electromagnetic fields as well as the number of fermions with right-left chirality \cite{Nielsen1981}.

Thus, from this perspective, one might wonder whether the inclusion of terms causing the violation of Lorentz
symmetry would also lead to interesting results, considering this relationship with condensed matter. In this sense,
one can apply QED to a class of materials that can be considered as Weyl semi-metals \cite{Balents2011, Wan2011, Balents22011, Balents32011, Rao2016}. With the
proper choice of physical parameters, these materials can be represented by the massive Dirac equation in (3 + 1)-dimensions (at low
energies, it is considered to describe quasi-particles), which is related to the following action (corresponding to only the $b_{\mu}$ term in equation (\ref{EQ1}))

\begin{equation}
S=\int d^4x \bar{\psi}(i \cancel{\partial}-m-\cancel{b}\gamma_{5}-e\cancel{A})\psi,
\label{Sbactionx}
\end{equation}
with $b_{\mu}$ being a constant four-vector.

This model has been studied with great enthusiasm, since it can generate a CFJ term \cite{CSI}-\cite{CSI11} that can be finite but undeterminated \cite{Jackiw}. However, starting from eq. (\ref{Sbactionx}), it is possible to radiatively induce the CFJ term and determine it unambiguously, since condensed matter models can be verified experimentally.


The corresponding action of eq. (\ref{Sbactionx}) for the condensed matter model in the momentum space is given by the expression
\begin{equation}
S=\int \frac{d^4 k}{(2\pi)^{4}} \bar{\psi}(\gamma_{\mu}M^{\mu}_{\nu}k^{\nu}-m-\cancel{b}\gamma_{5})\psi,
\label{Sbactionp}
\end{equation}
with $M^{\mu}_{\nu}=(v_{f},v_{f},\tilde{v}_{f})$ being a diagonal matrix which is necessary due to the anisotropy introduced by the Fermi velocity (in some materials, we 
consider $v_{f}=c/300$).


The amplitudes and propagators that come from eq. (\ref{Sbactionp}) are similar to the usual QED. The complete propagator is given by the expression
\begin{equation}\label{Prop1}
G(k,b)=\frac{i}{(\cancel{k}-m-\cancel{b}\gamma_{5})},
\end{equation}
with the polarization tensor $\Pi^{\mu\nu}$ given by the following expression, already adapted due to the velocity of propagation of the charge carriers on the graphene sheet being the Fermi velocity $v_{f}$,
\begin{equation}\label{Pimunu}
\Pi^{\mu\nu}(b,p)=\frac{e^{2}}{v_{f}\tilde{v}_{f}}\int \frac{d^4k}{(2\pi)^4}\ Tr[\gamma^{\mu}G(k,b)\gamma^{\nu}G(k+p,b)].
\end{equation}

Equation (\ref{Pimunu}) provides all the necessary information regarding the relationship between the Lorentz breaking and the Weyl semi-metals. It is now sufficient to perform a direct calculation to obtain information about the consequences of the parameter $b$. Thus, taking the trace and making the necessary calculations, we found the result for the amplitude given by equation (\ref{result}), in which only the $b_{\mu}$ parameter contributes to the result for conductivity. From the point of view of implicit regularization, which was discussed in detail in Section \ref{s2}, the parameter $\upsilon_{0}$, even when is required gauge (and/or momentum routing) invariance, is undetermined. In this sense, the process of fixing the parameter should be by phenomenology, like building an experimental apparatus that can measure the 4-current $j_{\mu}$ associated with the fermion current adapted to a condensed matter model (an example of measurement to fix such a parameter is an analysis of the Hall effect from the perspective of Weyl semi-metals, where it is possible to experimentally obtain the value of the so-called Hall conductivity \cite{Grushin}). On the other hand, from a theory point of view, some results agree that the parameter $\upsilon_{0}$ can be determined unambiguously for massless theories \cite{Perez1999,Brito2008}. 

After contracting the $A_{\mu}$ field with the finite piece that remains in eq. (\ref{Pi1}), we find out the following current

\begin{equation}\label{current1}
j^{\nu}=\frac{q^{2}}{2 \pi^{2}v_{f}\tilde{v}_{f}}(1-8\pi^{2}\upsilon_{0})b_{m}\epsilon^{m\nu\alpha\beta}\partial_{\alpha}A_{\beta},
\end{equation}
considering the spatial part and
\begin{equation}\label{current2}
j^{\nu}=\frac{q^{2}}{2 \pi^{2}v_{f}\tilde{v}_{f}}(1-8\pi^{2}\upsilon_{0})b_{0}\epsilon^{0\nu\alpha\beta}\partial_{\alpha}A_{\beta},
\end{equation}
when we consider the temporal part. The spatial part presented by (\ref{current1}) gives rise to the anomalous Hall effect, whose conductivity is proportional to the separation term between the Weyl nodes.
\begin{equation}
\sigma^{xy}=\frac{1}{2 \pi^{2}v_{f}\tilde{v}_{f}} \epsilon^{xyl}(1-8\pi^{2}\upsilon_{0})|\vec{b}| \hat{b}_{l}.
\end{equation}

The second equation, given by (\ref{current2}), describes the so-called magnetic effect, which sometimes implies the equilibrium of currents in the presence of the chiral 
magnetic field \cite{Zhou2013}. However, this is a naive statement, because the chiral anomaly in condensed matter models takes place near the Weyl nodes. Therefore, one must take a great care to predict chiral anomaly effects directly for a Weyl semi-metal (for more details on Weyl semi-metals, see \cite{Haldane2014}). Nevertheless, we see that Weyl semi-metals 
are interesting for studies in high energy physics, more specifically, considering that they can be analyzed from the point of view of Lorentz symmetry breaking, when the 
conductivity can be generated by a $\cancel{b}\gamma_{5}$ type term in the LV action.

A comment here is necessary. The $\Pi^{\mu\nu}$ polarization tensor modifies Maxwell equations. The even part is related to the characteristics of the electrical permeability 
and magnetic permeability constants of the medium where the electrons propagate. The odd part, on the other hand, can add new terms to Maxwell equations, which can modify the 
response of the material medium to the propagation of electrons in a Weyl semi-metal. Considering $j=\rho=0$ (absence of sources), the equation of the wave propagating in a Weyl 
semi-metal is modified, leading to the effect of vacuum birefringence associated with it. This effect is exclusively associated with the CFJ term induced radiatively and such 
observation is one example of how to fix the parameter $\upsilon_{0}$ by phenomenology. Some other interesting effects on Weyl semi-metals are described in \cite{Grushin2011} 
(Repulsive Casimir Effect) and \cite{Xi2009} (Axionic Electrodynamics) \footnote{A discussion about Weyl semi-metals theory and 
their relation with induced CFJ term and another models in high energy physics can be found in  \cite{Rao2016}}.

\section{Summary}
\label{s6}

We calculated the polarization tensor of the Abelian gauge field in a minimal LV extension of QED involving all terms listed in \cite{KosPic}. Within our studies, the main 
attention was paid, first, to divergent contributions, while most of previous studies dealt with finite ones, the most known of them is the CFJ term, second, to the 
infrared-leading parts of finite contributions. The importance of these terms is justified by the fact that they play a special role within condensed matter studies where the LV 
effects attract essential attention. In this study, we followed this line and calculated the anomalous Hall conductivity and discussed other possible applications of our results 
to the condensed matter, especially, to Weyl semi-metals.

A natural continuation of this study, besides of study of other applications of Lorentz symmetry breaking within the condensed matter context,  could consist in treating the 
low-energy impacts of higher-derivative LV terms. We are planning to do this study in a forthcoming paper.

\section*{Acknowledgments}

The work of A. Yu.\ P. has been partially supported by the CNPq project No. 301562/2019-9.

\section*{Appendix}

All integrals needed after taking the traces in eqs. (\ref{(a)})-(\ref{(f)}) can be obtained from integrals below:

\begin{align}
\centering
&\int_k \frac{1}{[(k-p)^2-m^2]}= I_{quad}(m^2)-p^2\upsilon_0;
\\
&\int_k \frac{k^{\alpha}}{[(k-p)^2-m^2]}= p^{\alpha}(I_{quad}(m^2)-\upsilon_2) -p^2p^{\alpha}(\xi_0-\upsilon_0);
\\
&\int_k \frac{k^{\alpha}}{[(k-p)^2-m^2]^2}= p^{\alpha}(I_{log}(m^2)-\upsilon_0);
\\
&\int_k \frac{k^{2}}{[(k-p)^2-m^2]^2}= I_{quad}(m^2)+(m^2+p^2)I_{log}(m^2) - 3 p^2 \upsilon_0;
\\
&\int_k \frac{k^{\alpha}k^{\beta}}{[(k-p)^2-m^2]^2}= \frac{1}{2}g^{\alpha\beta}(I_{quad}(m^2)- \upsilon_2)+p^{\alpha}p^{\beta}(I_{log}(m^2)-\xi_0)
-\frac{1}{2}p^2 g^{\alpha\beta}(\xi_0-\upsilon_0);
\\
&\int_k \frac{k^{2}k^{\alpha}}{[(k-p)^2-m^2]^2}= 2 p^{\alpha}(I_{quad}(m^2)- \upsilon_2)+p^{\alpha}(m^2+p^2)I_{log}(m^2)+p^{\alpha}(3 p^2-m^2)\upsilon_0-
4 p^2 p^{\alpha}\xi_0;
\\
&\int_k \frac{k^{\alpha}k^{\beta}k^{\gamma}}{[(k-p)^2-m^2]^2}= \frac{1}{2} p^{ \{ \alpha }g^{\gamma \beta \} }\left[I_{quad}(m^2)- \xi_2\right]
-\frac 12 p^2 p^{ \{ \alpha }g^{\gamma \beta \} } \left[ I_{log}(m^2) - \xi_0 \right] + \nonumber \\
& \;\;\;\;\;\;\;\;\;\;\;\;\;\;\;\;\;\;\;\;\;\;\;\;\;\;\;\;\;\;\;\; + \frac 12 (p^2 p^{ \{ \alpha }g^{\gamma \beta \} } + 2 p^\alpha p^\beta p^\gamma )\left[ I_{log}(m^2)-\sigma_0 \right]
;
\end{align}

\begin{align}
\centering
&I=\int_k \frac{1}{(k^2-m^2)^2[(k+p)^2-m^2]}= -b\ \int^1_0 dx \frac{(1-x)}{\Delta^2};
\\
&I_1^{\beta}=\int_k \frac{k^{\beta}}{(k^2-m^2)^2[(k+p)^2-m^2]}= b\ p^{\beta}\ \int^1_0 dx \frac{x(1-x)}{\Delta^2};
\\
&J_1^{\beta}=\int_k \frac{k^{\beta}}{(k^2-m^2)[(k-p)^2-m^2]^2}= I_1^{\beta}+p^{\beta}I;
\\
&I_2=\int_k \frac{k^2}{(k^2-m^2)^2[(k+p)^2-m^2]}= I_{log}(m^2)-b\ Z_0(p^2, m^2)-b\ m^2\ \int^1_0 dx \frac{(1-x)}{\Delta^2}
\\
&J_2=\int_k \frac{k^2}{(k^2-m^2)[(k-p)^2-m^2]^2}= I_2+p^2I+2p_{\beta}I_1^{\beta};
\\
&I_2^{\beta\nu}=\int_k \frac{k^{\beta}k^{\nu}}{(k^2-m^2)^2[(k+p)^2-m^2]}=\frac{1}{4}g^{\beta \nu}(I_{log}(m^2)-\upsilon_0 )
-\frac{1}{2} b\ g^{\beta \nu}[Z_0(p^2,m^2)-Z_1(p^2,m^2)]-\nonumber\\&-b\ p^{\beta}p^{\nu}\ \int^1_0 dx \frac{x^2(1-x)}{\Delta^2};
\\
&J_2^{\beta\nu}=\int_k \frac{k^{\beta}k^{\nu}}{(k^2-m^2)[(k-p)^2-m^2]^2}=I_2^{\beta\nu}+p^{\beta}p^{\nu}I+I_1^{\beta}p^{\nu}+I_1^{\nu}p^{
\beta};
\\
&I_3^{\nu}=\int_k \frac{k^2 k^{\nu}}{(k^2-m^2)^2[(k+p)^2-m^2]}= -\frac{1}{2} p^{\nu}(I_{log}(m^2)-\upsilon_0)+b\ p^{\nu}Z_1(p^2, m^2)+b\ m^2\ p^{\nu} \int^1_0 dx \frac{x(1-x)}{\Delta^2};
\\
&J_3^{\nu}=\int_k \frac{k^2 k^{\nu}}{(k^2-m^2)[(k-p)^2-m^2]^2}=I_3^{\nu}+p^{\nu}I_2+p^2 I_1^{\nu}+p^2 p^{\nu}I+ 2 p_{\gamma}I_2^{\gamma\nu}+ 2 p_{\gamma} p^{\nu
}I_1^{\gamma}-p^{\nu}\upsilon_0;
\\
&I_5^{\beta\nu\alpha}=\int_k \frac{k^{\beta}k^{\nu}k^{\alpha}}{(k^2-m^2)^2[(k+p)^2-m^2]}= \frac{-1}{12} b p^{\{ \alpha}g^{\beta \nu \}}(I_{log}(m^2)-\xi_0)+b\ p^{\alpha}p^{\beta}p^{\nu} \int^1_0 dx \frac{x^3(1-x)}{\Delta^2}+\nonumber\\
&+\frac{1}{2}b\ p^{\{ \alpha}g^{\beta \nu \}}[Z_1(p^2,m^2)-Z_2(p^2,m^2)];
\\
&J_5^{\beta\nu\alpha}=\int_k \frac{k^{\beta}k^{\nu}k^{\alpha}}{(k^2-m^2)[(k-p)^2-m^2]^2}= I_5^{\beta\nu\alpha}+p^{\beta}p^{\nu}p^{\alpha
}I+p^{\{ \nu}p^{\beta
}I^{\alpha \} }_1 +p^{\{ \beta}I^{\nu\alpha \}}_2 +\frac{1}{4}p^{\{ \alpha}g^{\beta\nu\}}(\upsilon_0-\xi_0)
\end{align}
\begin{align}
\centering
&I_4^{\beta\nu}=\int_k \frac{k^2 k^{\beta}k^{\nu}}{(k^2-m^2)^2[(k+p)^2-m^2]}= \frac{1}{2}g^{\beta \nu}(I_{quad}(m^2)-\upsilon_2)+\frac{1}{4}(m^2-p^2)g^{\beta \nu}(I_{log}(m^2)-\upsilon_0)+\nonumber\\
&+\frac{1}{6}(p^2 g^{\nu \beta}+2 p^{\beta} p^{\nu})(I_{log}(m^2)-\xi_0)- b\ (-g^{\beta \nu}p^2+p^{\nu}p^{\beta})Z_2(p^2,m^2)+\frac{1}{2}\ b(m^2-3p^2)\ g^{\beta \nu} Z_1(p^2,m^2)+\nonumber\\
&+\frac{1}{2}b(p^2-m^2)g^{\beta \nu}Z_0(p^2,m^2)-b\ p^{\beta}p^{\nu} m^2\ \int^1_0 dx \frac{x^2(1-x)}{\Delta^2};
\\
&J_4^{\beta\nu}=\int_k \frac{k^2 k^{\beta}k^{\nu}}{(k^2-m^2)[(k-p)^2-m^2]^2}= \frac{1}{2}g^{\nu\beta}(I_{quad}(m^2)-\upsilon_2)+
(m^2-p^2)J_2^{\beta\nu}+2p_{\lambda}J_5^{\beta\nu\lambda}-2p_{\lambda}I_5^{\beta\nu\lambda}-p^2 I_2^{\beta\nu},
\end{align}
where $b\equiv \frac{i}{(4\pi)^2}$, $Z_k(p^2,m^2)$ and $\Delta^2$ are defined as
\bq
Z_k(p^2,m^2)=\int^1_0 dz z^k \ln \frac{m^2-p^2 z(1-z)}{m^2},\\
\Delta^2=m^2-p^2 x(1-x).
\eq

The basic divergent integrals $I_{log}(m^2)$ and $I_{quad}(m^2)$ and the surface terms $\upsilon_0$, $\upsilon_2$ and $\xi_0$ are
defined in section \ref{s2}.

\end{document}